\documentclass[draft]{article}
\usepackage{amsmath}
\usepackage{amssymb}
\usepackage{graphicx}

\title{Pure quantum freezing of the 5$^{th}$ dimension}
\author{Vladimir Dzhunushaliev
\thanks{E-mail: dzhun@hotmail.kg}}

\begin{document}
\maketitle

\begin{center}
\textit{
Freie Universit\"at Berlin, Arnimallee 14, D-14195, Berlin, Germany\\
and \\
Dept. Phys. and Microel. Engineer., KRSU, Bishkek, \\
Kievskaya Str. 44, 720021, Kyrgyz Republic}
\end{center}

\begin{abstract}
It is shown that super thin and super long gravitational flux tube
solutions in the 5D Kaluza-Klein gravity have the Planck scale regions $(\approx
l_{Pl})$ where the metric signature changes from $\{ +,-,-,-,- \}$
to $\{ -,-,-,-,+ \}$. Such a change occurs too rapidly according to one of the
paradigms of quantum gravity which holds that the Planck length is
the minimal length in the nature and consequently the physical
quantities can not change very quickly over this length scale.
To avoid such a dynamic it is hypothesized that a pure quantum
freezing of the dynamics of the $5^{th}$ dimension takes place. As
a continuation of the flux tube metric in the longitudinal
direction the Reissner-Nordstr\"om metric is proposed. As a
consequence of such a construction one can avoid the appearance of a
point-like singularity in the extremal Reissner-Nordstr\"om
solution.
\end{abstract}
\date{}
\section{Introduction}

One of the intriguing problems of Kaluza-Klein gravity is how
the observed 4D universe can be connected with the hypothesized 5
(or multi) dimensional Kaluza-Klein gravity ? In this paper we
would like to show that on a super thin and  super long gravitational
flux tube there takes place a pure quantum freezing of the $5^{th}$ dimension.
This mechanism is based on one of paradigms of quantum gravity: 
\emph{the Planck length is the minimal length in the
nature}. It is easy to understand that this statement leads to the
conclusion that no physical quantity can vary on a
\textit{macroscopical} scale over the extent of such a \emph{microscopical}
length as the Planck length. We will use this statement for the
metric signature. For example, the following dynamic is impossible: having a
metric signature near $r < r_H - l_{Pl}$ as $\{ +,-,-,-,- \}$ and near
$r > r_H + l_{Pl}$ as $\{ -,-,-,-,+ \}$ (with $r_H$ is some constant).
According to the above statement in the region $|r - r_H|
< l_{Pl}$ there should be some quantum gravitational effects which will
prevent such a dynamic. In this paper we do not give any
microscopical description how this happens, but we consider only the
consequences of such a prohibition. Such an approach is similar to the
\emph{macroscopical} thermodynamical investigation of a physical
process when we are not interested in the \emph{microscopical}
description of the process.

\section{Short decription of super thin and super long gravitational flux tubes}

The 5D super thin and super long gravitational flux tube metric is
\begin{eqnarray}
ds^2 & = & \frac{dt^{2}}{\Delta(r)} - \Delta(r) e^{2\psi (r)}
\left [d\chi +  \omega (r)dt + Q \cos \theta d\varphi \right ]^2
\nonumber \\
&-& dr^{2} - a(r)(d\theta ^{2} +
\sin ^{2}\theta  d\varphi ^2),
\label{sec1-10}
\end{eqnarray}
where $\chi $ is the 5$^{th}$, extra coordinate; $r,\theta ,\varphi$
are $3D$ spherical-polar coordinates; $r \in \{ - \infty, + \infty \}$
is the longitudinal coordinate; $Q$ is the magnetic charge.
\par
The metric \eqref{sec1-10} gives us the following components of the
electromagnetic potential
$A_\mu$
\begin{equation}
A_t = \omega (r) \quad
\text{and} \quad
A_\varphi = Q \cos \theta
\label{sec1-20}
\end{equation}
and the Maxwell tensor
\begin{equation}
    F_{rt} = \omega' (r) \quad
    \text{and} \quad
    F_{\theta \varphi} = -Q \sin \theta .
\label{sec1-30}
\end{equation}
This means that we have radial Kaluza-Klein
electric $E_r \propto F_{tr}$ and magnetic
$H_r \propto F_{\theta \varphi}$ fields.
\par
Substituting this ansatz into the 5D Einstein vacuum equations
\begin{equation}
    R_{AB} - \frac{1}{2} \eta_{AB} R = 0
\label{sec1-35}
\end{equation}
$A,B = 0,1,2,3,5$ and $\eta_{AB}$ is the metric signature,
gives us
\begin{eqnarray}
    \frac{\Delta ''}{\Delta} - \frac{{\Delta '}^2}{\Delta^2} +
    \frac{\Delta 'a'}{\Delta a} + \frac{\Delta ' \psi '}{\Delta} +
    \frac{q^2}{a^2 \Delta ^2}e^{-4 \psi} & = & 0,
\label{sec1-40}\\
    \frac{a''}{a} + \frac{a'\psi '}{a} - \frac{2}{a} +
    \frac{Q^2}{a^2} \Delta e^{2\psi} & = & 0,
\label{sec1-50}\\
    \psi '' + {\psi '}^2 + \frac{a'\psi '}{a} -
    \frac{Q^2}{2a^2} \Delta e^{2\psi} & = & 0,
    \label{sec1-60}\\
    - \frac{{\Delta '}^2}{\Delta^2} + \frac{{a'}^2}{a^2} -
    2 \frac{\Delta ' \psi '}{\Delta} - \frac{4}{a} +
    4 \frac{a' \psi '}{a} +
    \frac{q^2}{a^2 \Delta ^2} e^{-4 \psi} +
    \frac{Q^2}{a^2} \Delta e^{2\psi} & = & 0
\label{sec1-70}
\end{eqnarray}
$q$ is the electric charge. These equations are derived after substituting
the expression
\eqref{sec1-80} for the electric field in the initial Einstein's equations.
The 5D $(\chi t)$-Einstein's equation (4D Maxwell
equation) is taken as having the following solution
\begin{equation}
  \omega ' = \frac{q}{a \Delta ^2} e^{-3 \psi} .
\label{sec1-80}
\end{equation}
For the determination of the physical sense of the constant $q$ let us
write the $(\chi t)$-Einstein's equation in the following way :
\begin{equation}
    \left( \omega ' \Delta ^2 e^{3\psi} 4 \pi a \right)' = 0.
\label{sec1-90}
\end{equation}
The 5D Kaluza - Klein gravity after the dimensional reduction indicates that
the Maxwell tensor is
\begin{equation}
    F_{\mu \nu} = \partial_\mu A_\nu - \partial _\nu A_\mu .
\label{sec1-100}
\end{equation}
That allows us to write the electric field as $E_r = \omega '$.
Eq.\eqref{sec1-90}, with the electric field defined by \eqref{sec1-100},
can be compared with the Maxwell's equations in a continuous medium
\begin{equation}
    \mathrm{div} \mathcal {\vec D} = 0
\label{sec1-110}
\end{equation}
where $\mathcal {\vec D} = \epsilon \vec E$ is an electric displacement
and $\epsilon$ is a dielectric permeability. Comparing Eq. \eqref{sec1-110}
with Eq. \eqref{sec1-90} we see that the magnitude
$q/a = \omega ' \Delta^2 e^{3\psi}$ is like to the electric displacement
and the dielectric permeability is $\epsilon = \Delta^2 e^{3\psi}$.
This means that $q$ can be taken as the Kaluza-Klein electric charge
because the flux of the electric field is
$\mathbf{\Phi} = 4\pi a\mathcal D = 4\pi q$.
\par
As the electric $q$ and magnetic $Q$ charges are
varied it was found \cite{dzhsin1} that the solutions to the metric in
Eq. \eqref{sec1-40}-\eqref{sec1-70} evolve in the following way :
\begin{enumerate}
\item
$0 \leq Q < q$. The solution is \emph{a regular gravitational flux tube}.
The solution is filled with both electric and magnetic fields. The longitudinal
distance between the $\pm r_H$ surfaces increases, and
the cross-sectional size does not increase as rapidly
as $r \rightarrow r_H$ with $q \rightarrow Q$. The values $r= \pm r_H$ are defined in the
following way $\Delta(\pm r_H)=0$.
\item
$Q = q$. In this case the solution is \emph{an infinite flux tube} filled
with constant electric and magnetic fields. The cross-sectional
size of this solution is constant ($ a= const.$).
\item
$0 \leq q < Q$. In this case we have
\emph{a singular gravitational flux tube}
located between two (+) and (-) electric and magnetic
charges at $\pm r_{sing}$. At
$r = \pm r_{sing}$ this solution has real singularities at the location of the charges.
\end{enumerate}
we focus on the case when $q \approx Q$ but $q > Q$. In this case
there is a region $|r| \leq r_H$ where the solution is similar to
a tube filled with almost equal electric and magnetic fields.
The length $L = 2 r_H$ of the throat of the flux tube (with $|r| < r_H$) depends on the 
relation $\delta = 1 - Q/q$, i.e.
$L \stackrel{\delta \rightarrow 0}{\longrightarrow} \infty$
but $\delta > 0$. A numerical analysis \cite{dzh3} shows that the
spatial cross section of the tube ($t, \chi , r = \mathrm{const}$) does not change
significantly. i.e. $a(r_H) \approx 2 a(0)$. The cross section of the tube
at the center $a(0)$ is arbitrary and we choose it as
$a(0) \approx l_{Pl}$. This gives a super thin and super long
gravitational flux tube: $L \rightarrow \infty$, $a(0) \approx l_{Pl}$.

\begin{figure}[h]
  \begin{center}
    \fbox{
    \includegraphics[height=7cm,width=7cm]{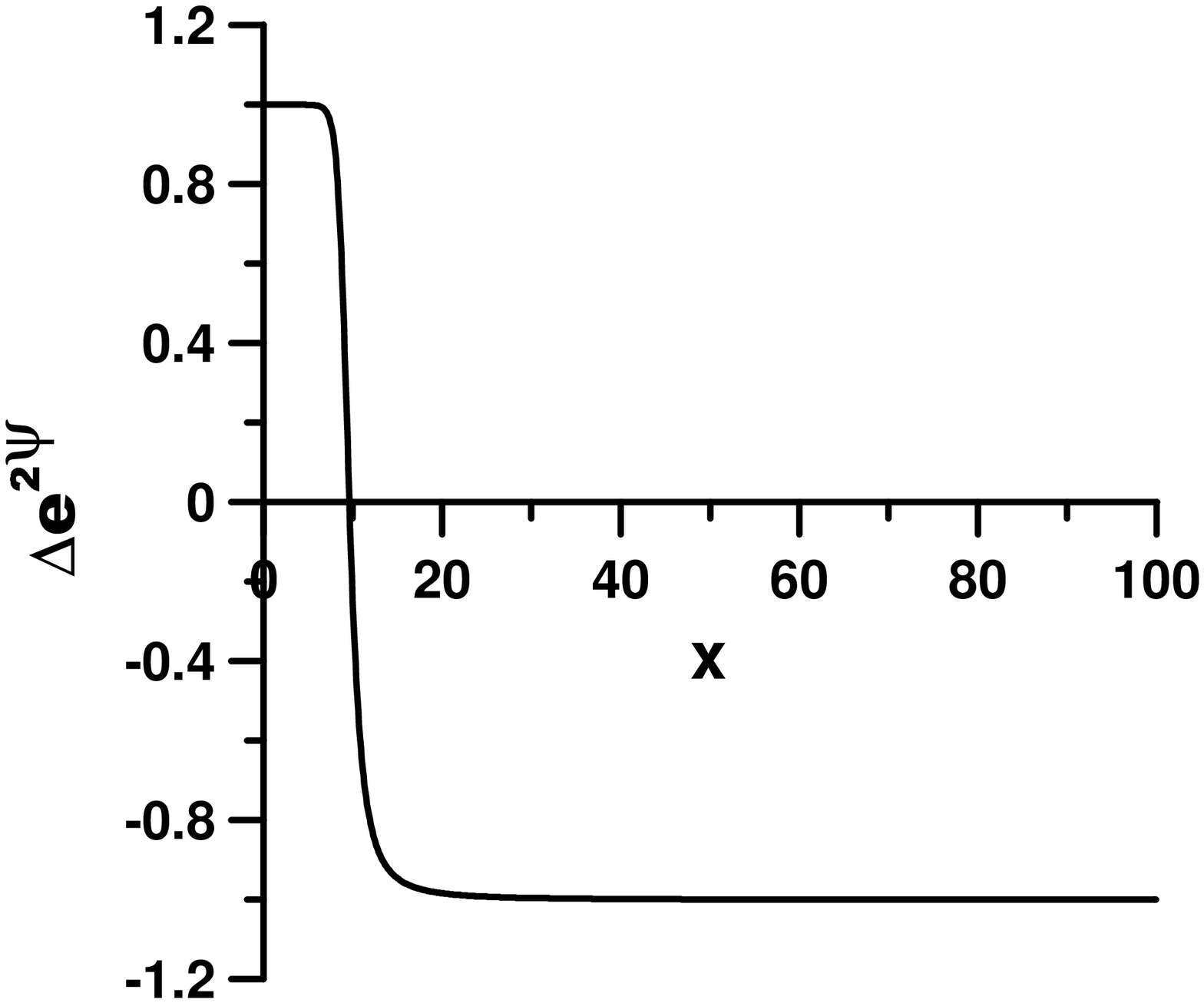}}
    \caption{The function $\Delta(x) e^{2\psi(x)}$.}
    \label{fig:prod}
  \end{center}
\end{figure}
\par
In Fig. \eqref{fig:prod} the profile of the function
$\Delta \mathrm e^{2\psi}$ (which is equal to $G_{55}$)
is presented \cite{dzh3}.
We see that near the values $r = \pm r_H$ this function
changes drastically from the value $\approx +1$ for
$-r_H +l_0 \lesssim r \lesssim r_H - l_0$ to
$\approx -1$ for $r \gtrsim r_H + l_0$ and $r \lesssim -r_H -l_0$.
The exact solution with $q = Q$ is
\begin{eqnarray}
    a &=& a(0) = \frac{Q_0^2}{2} = const,
\label{sec1-130}\\
    e^{2\psi} &=& \frac{1}{\Delta} = \cosh\frac{r}{\sqrt{a(0)}},
\label{sec1-140}\\
    \omega &=& \sqrt{2}\sinh\frac{r}{\sqrt{a(0)}} ,
\label{sec1-150}\\
        G_{55} & = & \Delta \mathrm e^{2 \psi} = 1
\label{sec1-155}
\end{eqnarray}
here we have parallel electric $E$ and magnetic $H$ fields with equal electric
$q_0$ and magnetic $Q_0$ charges $q_0 = Q_0 = \sqrt{2a(0)}$.
For the throat the solution with $q \approx Q$ but $q > Q$,
all the equalities of \eqref{sec1-130}-\eqref{sec1-155}
are changed to approximate equalities
\begin{eqnarray}
    a & \approx & \frac{Q_0^2}{2} = const,
\label{sec1-160}\\
    e^{2\psi} & \approx & \frac{1}{\Delta} = \cosh\frac{r}{\sqrt{a(0)}},
\label{sec1-170}\\
    \omega & \approx & \sqrt{2}\sinh\frac{r}{\sqrt{a(0)}} ,
\label{sec1-180}\\
    G_{55} & = & \Delta \mathrm e^{2 \psi} \approx 1
\label{sec1-190}
\end{eqnarray}
Such an approximation is valid only for $|r| \lesssim r_H - l_0$.
\par
Now we would like to estimate the length $l_0$ of the region
where the change of the function $\Delta \mathrm e^{2\psi}$ occurs
\begin{equation}
    \left. \Delta \mathrm e^{2\psi}\right|_{r \approx r_H + l_0} -
    \left. \Delta \mathrm e^{2\psi}\right|_{r \approx r_H - l_0} \approx 2 .
\label{sec1-195}
\end{equation}
For this estimation we use Eq. \eqref{sec1-50}.
On the throat this equation is approximately
\begin{equation}
    - \frac{2}{a} + \frac{Q^2}{a^2} \Delta e^{2\psi} \approx 0.
\label{sec1-200}
\end{equation}
We can estimate $l_0$ by solving Einsten's
equations \eqref{sec1-40}-\eqref{sec1-70} near $r = + r_H$
(for $r = - r_H$ the analysis is the same) and define $r = r_H - l_0$ where
the last two terms in Eq. \eqref{sec1-50} will have the same order
\begin{equation}
    \left.
        \left(
            \frac{2}{ a}
        \right)
    \right|_{r=r_H - l_0} \approx
    \left.
        \left(
            \frac{Q^2}{a^2} \Delta e^{2\psi}
        \right)
    \right|_{r=r_H - l_0}
\label{sec1-210}
\end{equation}
For the solution close to $r = \pm r_H$ we try the following form
\begin{equation}
    \Delta(r) = \Delta_1 \left( r_H - r \right) +
    \Delta_2 \left( r_H - r \right)^2 + \cdots .
\label{sec1-230}
\end{equation}
Substitution into Eq. \eqref{sec1-40} gives us the following
solution
\begin{equation}
    \Delta_1 = \frac{q e^{-2\psi_H}}{a_H}.
\label{sec1-260}
\end{equation}
After the substitution into Eq. \eqref{sec1-210} we have
\begin{equation}
    l_0 \approx \sqrt{a(0)} = l_{Pl}
\label{sec1-270}
\end{equation}
here we took into account that the numerical analysis of \cite{dzh3}
shows that $a_H \approx 2 a(0)$.
Thus the change of the macroscopical dimensionless function
$\Delta \mathrm e^{2\psi}$ as in Eq. \eqref{sec1-195} occurs
in the Planck length. The metric \eqref{sec1-10} near
$|r| \approx r_H - l_{Pl}$ is approximately
\begin{equation}
    ds^2 \approx \mathrm e^{2\psi_H} dt^2 - dr^2 - a(0)
    \left( d\theta^2 + \sin \theta d \varphi^2 \right) -
    \left(
        d\chi + \omega dt + Q \cos \theta d \phi
    \right)^2
\label{sec1-273}
\end{equation}
near $|r| \approx r_H + l_{Pl}$ the metric
is approximately
\begin{equation}
    ds^2 \approx -\mathrm e^{2\psi_H} dt^2 - dr^2 - a
    \left( d\theta^2 + \sin \theta d \varphi^2 \right) +
    \left(
        d\chi + \omega dt + Q \cos \theta d \phi
    \right)^2
\label{sec1-276}
\end{equation}
here we took into account that numerical calculations of \cite{dzh3} show
that $\psi \approx \psi_H = \mathrm {const}$ near $|r| \gtrsim r_H$.
We see that within the Planck length the metric signature changes
from $\{ +,-,-,-,- \}$ to $\{ -,-,-,-,+ \}$. Simultaneously it is necessary to
mention that the metric \eqref{sec1-10} is non-singular near $|r| = r_H$ and
approximately \cite{dzh3}
\begin{equation}
    ds^2 \approx
    g_H dt^2 -
    e^{\psi_H} dt \left( d\chi + Q \cos\theta d \varphi \right) -
    dr^2 - a(r_H) \left( d\theta^2 + \sin^2 \theta d\varphi^2 \right)
\label{sec1-280}
\end{equation}
where $g_H$ is some constant.
\par
If we write the metric \eqref{sec1-10} in 5-bein formalism
\begin{equation}
\begin{split}
    ds^2 &= \omega^A \omega^B \eta_{AB} ,\\
    \omega^A &= e^A_\mu dx^\mu , \quad
    x^\mu = t,r,\theta , \varphi , \chi
\label{sec1-290}
\end{split}
\end{equation}
then we see that
\begin{eqnarray}
  \eta_{AB} &=& \left\{ +1,-1,-1,-1,-1 \right\} \quad
  \text{by} \quad |r| \lesssim r_H - l_{Pl}
\label{sec1-300}\\
  \eta_{AB} &=& \left\{ -1,-1,-1,-1,+1 \right\} \quad
  \text{by} \quad |r| \gtrsim r_H + l_{Pl}
\label{sec1-310}
\end{eqnarray}
It is necessary to note that for the mechanism presented here the change
of the sign of two components $\eta_{00}$ and $\eta_{55}$ is very important.
The reason is that the $G_{55}$ metric component can be made dimensional in
the following way
\begin{equation}
    G_{55} d\chi ^2 = \left( l^2_0 G_{55} \right)\left( \frac{d \chi}{l_0} \right)^2
\label{sec1-320}
\end{equation}
where $l_0 \approx l_{Pl}$ is the characteristic length of the $5^{th}$
dimension. In this case the quantity $\sqrt{l_{Pl} G_{55}}$ changes
\begin{equation}
    \left. \sqrt{l_{Pl} G_{55}} \right|_{r \approx r_H + l_{Pl}} -
    \left. \sqrt{l_{Pl} G_{55}} \right|_{r \approx r_H - l_{Pl}}
    \approx l_{Pl}
\label{sec1-330}
\end{equation}
by a change of the radial coordinate
\begin{equation}
    \Delta r \approx l_{Pl}.
\label{sec1-340}
\end{equation}
Such a variation of $G_{55}$ is possible but simultaneously the dimensionless
quantity $\eta_{00} (e^0_t)^2$ changes
\begin{equation}
    \left. \eta_{00} (e^0_t)^2 \right|_{r \approx r_H + l_{Pl}} -
    \left. \eta_{00} (e^0_t)^2 \right|_{r \approx r_H - l_{Pl}}
    \approx 2 \mathrm e^{2\psi_H} \gg 1
\label{sec1-350}
\end{equation}
within a Planck length.
\par
One of the basic paradigms of quantum gravity is that the Planck
length is the minimal length in nature and consequently no
physical fields or quantities can change in the course of the
Planck length. Thus one can conclude that such a classical dynamic
of the metric signature is
\emph{impossible}. There is only one way to avoid such dynamical behavior of
this quantity: it is necessary to forbid \emph{any} dynamic
of the $G_{55}$ field variable so that its last value is conserved, or in
other words one has \emph{a pure quantum freezing of the dynamic of $5^{th}$
dimension}. Mathematically this means that $G_{55}$ becomes a non-dynamical
quantity and must not vary.
\par
Thus the analysis of the classical dynamic of
the metric signature for the super thin and super long gravitational
flux tube shows that there
is a region where this quantity changes too quickly from the quantum
gravity viewpoint. This leads to the fact that some pure quantum gravity
effects have to happen in order to avoid such a variation of
the metric signature. We do not
consider the mechanism of these effects but the author's point of view
is that such mechanism can not be based on any field-theoretical consideration.
This pure quantum freezing of the extra dimension is similar to a trigger which
has only two states: in one state the dynamic of $G_{55}$ is switched
on, and in another is switched off. Such a quantum dynamic,
which can be realized only in the Planck region, is a non-differentiable
dynamic. Examples of such a hypothesized non-differentiable dynamic
are the above mentioned freezing of the extra dimensions, the change of the metric
signature, or other phenomena \cite{Bogdanoff:2001dz}, \cite{castro}.
\par
The next question arising in this context is the dynamic of residual
degrees of freedom. Since $G_{55} = \mathrm{const}$ we have Kaluza-Klein
gravity in its initial interpretation where 5D Kaluza-Klein theory with
$G_{55} = \mathrm{const}$ is equivalent to 4D electrogravity. It lead to an
idea that the spacetime with ($r \gtrsim r_H + l_{Pl}$)
(the same for $r \lesssim -r_H - l_{Pl}$) will be the Reissner-Nordstr\"om
solution with the corresponding electric and magnetic fields.

\section{Joining of electric and magnetic fields}

In this section we would like to discuss the problem of joining
the electric and magnetic fields of the flux tube and the Reissner-Nordstr\"om
solutions. For this we will compare the fluxes of electric and magnetic
fields on the throat and the Reissner-Nordstr\"om spacetime.

\subsection{Electric field}

At first we consider the electric field. The Maxwell equations for the 4D case are
\begin{equation}
    \frac{1}{\sqrt{\gamma}} \frac{\partial}{\partial x^\alpha}
    \left(
        \sqrt{\gamma} \stackrel{(4)}{D}\!\!{}^\alpha
    \right) = 0
\label{sec2-10}
\end{equation}
where $\sqrt{\gamma}$ is the determinant of the 3D spatial metric and
$\stackrel{(4)}{D}\!\!{}^\alpha$ is an analog of the electric displacement in the
electrodynamics of a medium and
\begin{equation}
    \stackrel{(4)}{D}\!\!{}^\alpha = - \sqrt{g_{00}} \stackrel{(4)}{F}\!\!{}^{0 \alpha} .
\label{sec2-20}
\end{equation}
For the Reissner-Nordstr\"om solution the metric is
\begin{equation}
\begin{split}
    \stackrel{(4)}{ds}\!\!{}^2 = &\left( 1 - \frac{r_g}{R} + \frac{r_{q,Q}}{R^2} \right)
    \stackrel{(4)}{dt}\!\!{}^2 +
    \frac{dR^2}{1 - \frac{r_g}{R} + \frac{r_{q,Q}}{R^2}} -
    R^2 \left( d \theta^2 + \cos^2 \theta d \varphi^2 \right),\\
    & r_g = \frac{2 G m}{c^2}, \quad
    r_{q,Q} = \sqrt{\frac{G}{c^4} \left( q^2 + Q^2 \right)}.
\end{split}
\label{sec2-30}
\end{equation}
Therefore Eq. \eqref{sec2-10} is
\begin{equation}
    \frac{d}{d R}
    \left(
        R^2 \stackrel{(4)}{F}\!\!\!{}_{tR}
    \right) = 0 , \quad
    \stackrel{(4)}{F}\!\!{}_{tR} = \frac{d \phi}{dR}
\label{sec2-40}
\end{equation}
here $\phi$ is the scalar potential. This equation shows that the flux
$\Phi_{e}$ of the electric field $E_R = \phi '$ is conserved
\begin{equation}
    \stackrel{(4)}{\Phi}\!\!{}_{e} = 4 \pi R^2 \phi ' = 4 \pi \!\!\stackrel{(4)}{q} =
    \mathrm{const} .
\label{sec2-50}
\end{equation}
The corresponding Maxwell equation on the throat is
\begin{equation}
     \frac{d}{d r}
     \left[
        a \left(
                \omega ' \Delta^2 \mathrm{e}^{3 \psi}
            \right)
     \right] = 0.
\label{sec2-60}
\end{equation}
Here we have also a conserved flux of an analog of electric displacement
\begin{equation}
    \stackrel{(5)}{\Phi}\!\!{}_{e} = 4 \pi a \!\!\stackrel{(5)}{D} \!\!{}^r =
    4 \pi \!\!\stackrel{(5)}{q} = \mathrm{const}
\label{sec2-70}
\end{equation}
where
\begin{equation}
    \stackrel{(5)}{D} \!\!{}^r = \omega ' \Delta^2 \mathrm{e}^{3 \psi} =
    \omega ' \mathrm{e}^{-\psi} G_{55}.
\label{sec2-73}
\end{equation}
The simplest assumption about two electric fields on the throat and
Reissner-Nordstr\"om spacetimes is to join the fluxes \eqref{sec2-50} and
\eqref{sec2-70}
\begin{equation}
    \stackrel{(5)}{\Phi}\!\!{}_{e} = \stackrel{(4)}{\Phi}\!\!{}_{e}
\label{sec2-80}
\end{equation}
here $\stackrel{(5)}{\Phi}\!\!\!\!{}_{e}$ have to be calculated at the ends of throat where
the freezing of the $5^{th}$ coordinate happens, i.e. near $r \approx r_H - l_{Pl}$
and $r \approx -r_H + l_{Pl}$. There $G_{55} \approx 1$ and
\begin{eqnarray}
    \phi '& \approx & \omega '\mathrm{e}^{-\psi_H}
\label{sec2-90}\\
    \stackrel{(5)}{q} & = & \stackrel{(4)}{q} = q
\label{sec2-100}
\end{eqnarray}
here we have used the conditions
\begin{eqnarray}
    \left. \stackrel{(4)}{g}\!\!\!{}_{\theta \theta} \right|_{R = r_0} & = & r_0^2 =
    a(0) = \left. \stackrel{(5)}{G}\!\!{}_{\theta \theta} \right|_{r = r_H - l_{Pl}}
\label{sec2-110}\\
    \psi_H & \approx & \left. \psi \right|_{r = r_H - l_{Pl}}
\label{sec2-120}
\end{eqnarray}

\subsection{Magnetic field}

Now we will repeat a similar calculation for the magnetic field. The Maxwell
equation for the magnetic field having the information about the flux of magnetic
field is the same for the throat and the tails (Reissner-Nordstr\"om spacetimes)
\begin{equation}
    \frac{1}{\sqrt{\gamma}} \frac{\partial}{\partial x^\alpha}
    \left[
        \sqrt{\gamma} \left(
            \frac{-1}{2\sqrt{\gamma}} \epsilon^{\alpha\beta\delta} F_{\beta\delta}
        \right)
    \right] = 0
\label{sec2b-10}
\end{equation}
here the magnetic field $B^\alpha$ can be introduced
\begin{equation}
    B^\alpha = \frac{-1}{2\sqrt{\gamma}} \epsilon^{\alpha\beta\delta} F_{\beta\delta}.
\label{sec2b-20}
\end{equation}
More concretely
\begin{equation}
    \frac{d}{dr} \left[
        r^2 \left(
            \frac{Q}{r^2}
        \right)
    \right] = 0
\label{sec2b-30}
\end{equation}
here $r$ is the radial coordinate on the throat and the tails. It allows us to
introduce the flux of magnetic field for the throat and the tails which will be equal
\begin{equation}
    \stackrel{(4)}{\Phi}\!\!{}_m = 4 \pi R^2 \frac{\stackrel{(4)}{Q}}{R^2} =
    4 \pi a \frac{\stackrel{(5)}{Q}}{a} = \stackrel{(5)}{\Phi}\!\!{}_m .
\label{sec2b-40}
\end{equation}
This permits us to introduce the magnetic charge
\begin{equation}
    \stackrel{(4)}{Q} = \stackrel{(5)}{Q} = Q .
\label{sec2b-50}
\end{equation}

\subsection{Joining the metric}

After the quantum freezing of the $5^{th}$ dimension the 5D metric \eqref{sec1-10} will be
\begin{equation}
    \stackrel{(5)}{ds}\!\!{}^2 = A(r)dt^{2} - B(r) dr^{2} - a(r)(d\theta ^{2} +
    \sin ^{2}\theta  d\varphi ^2) -
    \left [d\chi +  \omega (r)dt + Q \cos \theta d\varphi \right ]^2 .
\label{sec2c-10}
\end{equation}
The interpretation of this metric in the initial Kaluza-Klein gravity tells us
that we have the 4D metric
\begin{equation}
    \stackrel{(4)}{ds}\!\!{}^2 = A(r)dt^{2} - B(r) dr^{2} - a(r)(d\theta ^{2} +
    \sin ^{2}\theta  d\varphi ^2)
\label{sec2c-20}
\end{equation}
and the electromagnetic potential
\begin{equation}
    A_\mu = \left\{ \omega ,0,0, Q \cos \theta \right\}.
\label{sec2c-30}
\end{equation}
In the previous subsections we have joined the electric and magnetic fields.
Now we have to consider the metric components. We will interpret the metric
\eqref{sec2c-20} as a Reissner-Nordstr\"om metric. After freezing of the $5^{th}$
coordinate the corresponding equations become 5D Einstein's equations but
without the $(55)$ Einstein equation, i.e. we have the ordinary 4D electro-gravity.
Taking this into account the metric \eqref{sec2c-10} will be the Reissner-Nordstr\"om
metric \eqref{sec2-30}. We will join the $g_{\theta\theta}$ and $g_{tt}$ components
of the metrics \eqref{sec1-10} and \eqref{sec2c-20}. One can note that joining
the $g_{rr}$ components of these metrics is not necessary as they measure
the distance on the transversal direction to the surface of joining.
We do not join (as is normally done) the first derivatives of the metric components since
on the 5D throat and the tails there are \textit{the different sets of equations}: 
on the throat one has the 5D Einstein's equations but on the tails one h
as the equations of 4D electro-gravity
\footnote{Let us
remember that according to the initial interpretation of the Kaluza-Klein
gravity the 4D electro-gravity spacetime can be considered as 5D spacetime
with frozen $5^{th}$ dimension ($G_{55} = \mathrm{const}$)}.
\par
Joining of $g_{\theta \theta}$ gives
\begin{equation}
    \stackrel{(4)}{g}\!\!{}_{rr} \left( r_0 \right) =
    \left. a \right|_{\pm r_H \mp l_{Pl}} \quad
    \Rightarrow \quad r_0^2 = a(0) = l_{Pl}.
\label{sec2c-40}
\end{equation}
For the $g_{tt}$ components
\begin{equation}
    \left(
        1 - \frac{r_g}{r_0} + \frac{r^2_{q,Q}}{r_0^2}
    \right)
    \stackrel{(4)}{dt}\!\!{}^2 = \mathrm{e}^{2\psi_H}\stackrel{(5)}{dt}\!\!{}^2
\label{sec2c-50}
\end{equation}
here $\stackrel{(5)}{t}\!\!{}^2$ and $\stackrel{(4)}{t}\!\!{}^2$ are the time coordinates
on the throat and the tails correspondingly. One can rewrite this relation as
\begin{equation}
    \frac{\stackrel{(5)}{dt}}{\stackrel{(4)}{dt}} =
    \mathrm{e}^{-\psi_H}
    \sqrt{1 - \frac{r_g}{l_{Pl}} + \frac{r^2_{q,Q}}{l_{Pl}^2}}.
\label{sec2c-60}
\end{equation}
Only with such a relation between the time coordinates on the throat and the tails
will time pass equally on the hypersurface of the junction.
\par
It is necessary to mention that the analysis presented in this section is very
simple and, for example, does not allow us to determine the mass
$m$ for the Reissner-Nordstr\"om solution. A more exact calculation
might be possible using quantum field-theoretical language.

\section{Conclusions and discussion}

In this paper we have shown that the super thin and super long
gravitational flux tube solutions in 5D Kaluza-Klein gravity has
two regions where the classical description can not be applied. Some
metric components change too quickly: the metric signature changes
from $\{ +,-,-,-,- \}$ to $\{ -,-,-,-,+ \}$ and $\Delta G_{55}
\approx 2$ within a Planck length. To avoid such a variation the
dynamical quantity $G_{55}$ must be frozen. Then the initial flux
tube metric can be extended on the right and left ends to the
Reissner-Nordstr\"om spacetimes \footnote{Let us note that additionally one
can freeze 4D metric $g_{\mu \nu}$ and we will have only Maxwell 
electrodynamic on the flat space which is similar to the AdS/CFT
correspondence idea.}. At the junctions there occurs \emph{a pure
quantum freezing of the dynamic of $G_{55}$ metric component}. Such
an object looks like two extremal Reissner-Nordstr\"om spacetimes
($r^2_g/4 < r^2_{q,Q}$) connected with a super thin and super long
flux tube filled with the electric and magnetic fields. Let us note
that the point of view presented here is based only on an idea
which does not depend on the details whatever theory of quantum
gravity one considers.
\par
Such a model allows us to resolve successfully one of the hardest problems of
general relativity: the avoidance of singularities. Our construction shows that
for an extremal Reissner-Nordstr\"om solution (at least for $q > Q$) the gravitational
field becomes so strong that the dynamic on the extra dimension becomes excited
and a flux tube from one singularity to another one appears. The force lines do not
convergence in a pointlike singularity but leave our universe to another one
(or to a remote part of our universe) through the flux tube and there appear
near another almost singularity.
\par
There is another argument for the pure quantum freezing mechanism presented here.
We see that near $|r| \lesssim r_H - l_{Pl}$ the metric component
$G_{55} \approx 1$ to a large accuracy.
This means that with the same accuracy we have a \emph{dynamical} freezing of the
$5^{th}$ coordinate but near the surface where the change of metric signature
occurs the quantity $G_{55}$ becomes dynamical. 
\par
Let us note that such mechanism works only for some special extremal
Reissner-Nordstr\"om solutions having electric and magnetic charges with
the relation $q > Q$. The question arises: is it possible to extend
this result about avoidance of a singularity to other Reissner-Nordstr\"om solutions ?
In this connection we have to emphasize that the analysis carried out here
is very simple and a more careful analysis with the quantization of
corresponding field quantities can give more exact results. In this connection one can
mention the results of Ref. \cite{Dzhunushaliev:2004wj} where it is shown that the super thin
and super long flux tube solutions without freezing of the $5^{th}$ dimension have interesting
properties on the tails: the magnetic fields decreases faster then the electric
field and there is a rotation connected with the magnetic field on the throat.
One can hypothesis that these properties will remain the same with the
freezing of the $5^{th}$ coordinate (maybe in some weaker form).
\par
The construction presented here (two Reissner-Nordstr\"om spacetimes connected by
a flux tube) can be considered as a model of a spacetime with a frozen $5^{th}$
dimension but with piecewise sections where the $5^{th}$ dimension is unfrozen.
Similar ideas about spacetime regions with compactified and uncompactified
phases was considered by Guendelman in Ref. \cite{Guendelman:1991pc}. The 
construction presented in \cite{Guendelman:1991pc} 
is also based on a flux tube (i.e. the Levi-Civita - Bertotti - Robison
solution \cite{levi-civita}-\cite{robinson}) inserted between two 4D black holes.
Similarly in Ref. \cite{zaslavski} a model of the classical electron with 
a Levi-Civita -Bertotti -Robison flux tube between two Reissnaer-Nordstr\"om 
black holes is considered. 

\section{Acknowledgment}

I am very grateful to the Alexander von Humboldt Foundation for the financial support
and thanks Prof. H. Kleinert for hospitality in his research group and 
D Singleton for the fruitful discussion.


\begin{thebibliography}{99}

\bibitem{dzhsin1}
V. Dzhunushaliev and D. Singleton, Phys. Rev. \textbf{D59},
064018 (1999).

\bibitem{dzh3}
V. Dzhunushaliev, ``Strings and Branes under Microscope'', gr-qc/0312038;
''Wormhole solutions in 5D Kaluza-Klein theory as string-like objects´´,
gr-qc/0405017, to be published in ``Progress in General Relativity and 
Quantum Cosmology Research''.

\bibitem{Bogdanoff:2001dz}
G.~Bogdanoff and I.~Bogdanoff,
Class.\ Quant.\ Grav.\,  {\bf 18}, 4341(2001);
V.~Dzhunushaliev and D.~Singleton,
Class.\ Quant.\ Grav.\,  {\bf 18}, 1787(2001).

\bibitem{castro}
C.~Castro and M.~Pavsic,
Int.\ J.\ Theor.\ Phys.\,  \textbf{42}, 1693 (2003).

\bibitem{Dzhunushaliev:2004wj}
V.~Dzhunushaliev,
Phys. Lett. \textbf{B600}, 171 (2004).

\bibitem{Guendelman:1991pc}
E.~I.~Guendelman,
Gen.\ Rel.\ Grav.\  {\bf 23} (1991) 1415.

\bibitem{levi-civita}
T. Levi-Civita,
Atti Acad. Naz. Lincei, \textbf{26}, 519 (1917).

\bibitem{bertotti}
B. Bertotti,
Phys. Rev., \textbf{116}, 1331 (1959).

\bibitem{robinson}
I. Robinson,
Bull. Akad. Pol., \textbf{7}, 351 (1959).

\bibitem{zaslavski}
O.~B.~Zaslavskii,
Phys.\ Rev.\ D, \textbf{70}, 104017 (2004).



\end{thebibliography}
\end{document}